\documentstyle[preprint,aps]{revtex}
\tightenlines

\begin{document}

\title{Reply to Comment on ``Proposal for the Measurement of 
Bell-Type Correlations from Continuous Variables''}
\author{T. C. Ralph and W. J. Munro}
\address{Department of Physics, Centre for Laser
Science,\\ University of Queensland, QLD 4072, Australia
\\Fax: +61 7 3365 1242  Telephone: +61 7 3365 3412 \\ 
email: ralph@physics.uq.edu.au\\}
\maketitle


\begin{abstract}

Recently K. Banaszek, I. A. Walmsley, K. Wodkiewicz 
(quant-ph/0012097) commented on our Proposal for the Measurement of 
Bell-Type Correlations from Continuous Variables [T. C. Ralph, W. J. Munro, 
R. E. S. Polkinghorne, Phys. Rev. Lett. 85, 2035 (2000)]. Their 
comment is based on a blatant misreading and misunderstanding of our 
letter and as such is simply wrong.

\end{abstract}

\vspace{5mm}

The comment by K. Banaszek, I. A. Walmsley, K. Wodkiewicz \cite{banana} 
on our letter \cite{ral} purports to present a hidden variable theory 
which describes our results.  The hidden variable scheme described is 
based on the positive Wigner distribution which describes quadrature 
measurements on the parametric amplifier and vacuum background.  It is 
well known, and was pointed out on three separate occasions in our 
paper, that such a description is possible if {\it only} 
quadrature measurements are performed.  This is the reason why (as was 
also pointed out on three separate occasions) auxiliary {\it intensity} 
measurements must also be made.  The comment 
incorrectly identifies these auxiliary measurements as being 
quadrature measurements of the vacuum.  Instead, as is stated 
in the paper, they are quantum limited intensity measurements on the 
vacuum. 
 
As is also clearly stated in the paper, 
this auxiliary intensity 
measurement guarantees the positivity of the underlying (but 
inaccessible) individual measurements which make up our ensemble average. 
Without such a measurement the positivity of the individuals is not 
guaranteed and the Bell test cannot be applied. The authors say 
nothing that was not already stated in our letter when they make this 
point in their comment. 

It seems a cursory reading of our letter can lead to misconceptions. 
Given the brevity of the letter format this is perhaps understandable. 
A more detailed account is presently in preparation. Here we present a 
brief discussion aimed particularly at the point which the comment 
missed.

The crux of the issue is whether the positivity of our local realities, the count 
rates $R^{i}_{A}(\theta_{A})$ and $R^{j}_{B}(\theta_{B})$, given by
\begin{eqnarray}
R_{A}^{i}(\theta_{A}) & = & 
(\hat X_{A;1}^{i})^{2}-(\hat X_{va;1}^{i})^{2}+
(\hat X_{A;2}^{i})^{2}-(\hat X_{va;1}^{i})^{2}\nonumber\\
R_{B}^{j}(\theta_{B}) & = & 
(\hat X_{B;1}^{j})^{2}-(\hat X_{vb;1}^{j})^{2}+
(\hat X_{B;2}^{j})^{2}-(\hat X_{vb;1}^{j})^{2}
\label{r}
\end{eqnarray}
can be guaranteed. Here as usual the quadrature operators are related 
to their corresponding annihilation ($\hat C$) and 
creation ($\hat C^{\dagger}$) operators via
\begin{eqnarray}
\hat X_{C;1} & = & \hat C^{\dagger}+\hat C \nonumber\\
\hat X_{C;2} & = & i(\hat C^{\dagger}-\hat C) \nonumber\\
\hat C & = & \hat A, \hat B, \hat V_{A}, \hat V_{B}
\label{r2}
\end{eqnarray}
where $\hat A$ and $\hat B$ represent the polarization entangled 
beams at stations $A$ and $B$, whilst $\hat V_{A}$ and $\hat V_{B}$ 
represent vacuum modes which enter the detectors when the entangled 
beams are blocked. The problem is that the different quadrature 
components cannot be measured simultaneously (just as results for 
different polarization angles cannot be measured simultaneously). 
Thus, although the ensemble averages are positive, we cannot guarantee 
the positivity of individual realizations $R^{i}_{A}(\theta_{A})$ or 
$R^{j}_{B}(\theta_{B})$ from separate quadrature measurements. 
However, by substituting Eq.\ref{r2} into Eq.\ref{r}, 
it is straightforward to prove the equalities
\begin{eqnarray}
R_{A}^{i}(\theta_{A}) & = & 
\hat {A^{i}}^{\dagger} \hat A^{i}-
\hat {V^{i}_{A}}^{\dagger} \hat V^{i}_{A}\nonumber\\
R_{B}^{j}(\theta_{B}) & = & 
\hat {B^{j}}^{\dagger} \hat B^{j}-
\hat {V^{j}_{B}}^{\dagger} \hat V^{j}_{B}
\label{r1}
\end{eqnarray}
It is very important to note that these equalities hold for both a 
quantum {\it and a classical} treatment of the observables. That is 
Eq.\ref{r1} is equally true whether the variables are treated as 
operators or c-numbers. It is 
then trivial to see from Eq.\ref{r1} that provided $\hat V_{A,B}$ 
represents an unoccupied mode 
(ie is in a vacuum state) then the positivity of our local realities 
is guaranteed. In other words to guarantee the positivity of ${R_{A}}^i$ we 
must show explicitly through measurement that $V^{\dagger} V=0$. 
This is what our auxiliary measurement does. Thus the measurement of the 
intensity of the vacuum is a necessary but sufficient condition for 
the positivity of the local realities and hence the validity of the 
Bell inequality for the quadrature ensemble averages.

The auxiliary measurement could be included in the measurement protocol 
by randomly swapping between it and the various quadrature 
measurements, thus avoiding conspiracy loopholes. 
For sufficiently large data sets no information would 
need to be discarded. Indeed in principle the auxiliary measurements 
could be made with the same detectors as the quadrature measurements by 
simply blocking both the local oscillator and the signal at various 
times. In practice, though, detectors with sufficient dynamic range to make 
this possible do not presently exist.

In summary the comment is based on an erroneous assumption: the incorrect 
identification of the auxiliary measurement in our letter as a quadrature 
measurement. As a result it fails to raise any issues which were not in fact 
discussed in our letter.


\end{document}